 \documentclass[twocolumn]{aastex6}
\usepackage{amsmath}
\usepackage{amssymb}
\usepackage{amsfonts}
\usepackage{orcidlink} 
\usepackage[T1]{fontenc} 

\usepackage{color}
\usepackage{multirow}
\usepackage{amsmath}

\newcommand{\bwt}{\begin{widetext}}
\newcommand{\ewt}{\end{widetext}}
\newcommand{\beq}{\begin{equation}}
\newcommand{\eeq}{\end{equation}}
\newcommand{\bea}{\begin{eqnarray}}
\newcommand{\eea}{\end{eqnarray}}

\begin{document}
                    
\title{Sub-stellar Strange Quark Matter Objects: Predicting a New Class of Highly-Compact Candidates}

\author{Jonathan Jo\'as Zapata Campos \orcidlink{0000-0002-8577-3226}} 
\affiliation{Universidad Tecnol\'ogica del Per\'u, Lima, Per\'u}
\email{jzapata@utp.edu.pe} 

\author{Rodrigo Negreiros \orcidlink{0000-0002-9669-905X}}
\affiliation{Department of Physics, Catholic Institute of Technology, MA, USA}
\affiliation{Department of Physics, Universidade Federal Fluminense, Niteroi, Brazil}
\affiliation{ICRANet, Piazza della Repubblica 10, I-65122 Pescara, Italy}
\email{rnegreiros@catholic.tech}


\begin{abstract}
We investigate the existence and stability of highly-compact sub-stellar objects composed of strange quark matter (SQM), focusing on finite-size strangelets with baryon number $A \leq 100$. Motivated by the emergence of mass--radius outliers in the \textit{Gaia} DR3 era, we employ a Bayesian exploration of the MIT bag-model parameter space, explicitly accounting for finite-size surface and curvature contributions that become relevant at low baryon number. Enforcing the bulk absolute-stability requirement for SQM ($E/A < 930~\mathrm{MeV}$), we find that self-gravitating equilibrium sequences are confined to the sub-stellar regime, with typical masses $M \simeq 10^{-2}$--$10^{-1}\,M_{\odot}$ and characteristic radii of order $10^{3}$--$10^{4}$ km. We further show that 
rapid rotation, treated through a self-consistent framework that incorporates relativistic thermodynamics, 
can substantially inflate the equatorial radius and extend the accessible mass--radius domain. While rotation does not eliminate the intrinsic high-density compactness of these configurations, it shifts the most extended models closer to the observational parameter space of massive exoplanets. A comparison with objects from the NASA Exoplanet Archive reveals a pronounced density gap separating standard atomic-matter planets and brown dwarfs from the strangelet-rich branch predicted here. We conclude that light strangelets cannot account for solar-mass white dwarfs, but they robustly predict a previously unexplored population of ultra-compact sub-stellar objects, offering testable targets for future microlensing searches and high-cadence photometric surveys.
\end{abstract}


\vspace*{5mm}
\section{Introduction}\label{sec:Int}
\label{sec:intro}
White dwarfs (WDs) represent the final evolutionary stage of the majority of main-sequence stars and constitute high-density astrophysical laboratories for testing matter under extreme conditions \citep{Chandrasekhar1931}. 
The standard picture---carbon--oxygen (C/O) cores supported by electron degeneracy pressure---successfully explains the observed WD mass--radius ($M$--$R$) relation over a wide range of masses. 
However, the advent of high-precision astrometry from the \textit{Gaia} mission \citep{GaiaDR3_2023} has highlighted a subset of compact candidates reported as unusually massive and/or unusually compact compared with classical expectations \citep{Althaus2022,Kilic2021}. 
While the interpretation of these outliers is still under active scrutiny, they motivate renewed interest in whether non-standard compositions or exotic phases of matter could produce distinct equilibrium branches beyond the canonical C/O sequence.

A compelling possibility is offered by the Bodmer--Witten--Terazawa hypothesis, which posits that strange quark matter (SQM)---a roughly charge-neutral mixture of up, down, and strange quarks---may represent the true ground state of hadronic matter \citep{Bodmer1971, Witten1984}. 
Self-bound SQM configurations are commonly discussed as strange stars \citep{Weber2005}, whereas hybrid objects featuring ordinary-matter crusts or clustered strangelet phases have been explored under the broader category of strange dwarfs \citep{Glendenning1995a, Glendenning1995b}. 
Most existing studies have focused on macroscopic SQM phases or on strangelet clusters in the large-baryon-number regime ($A \gtrsim 10^{3}$), where finite-size corrections are subdominant. 
By contrast, the microphysical viability and structural impact of \emph{light} strangelets ($A \leq 100$) remain comparatively less understood, precisely in the regime where surface and curvature contributions can control stability and the resulting equation of state (EOS) \citep{Madsen1993, Alford2012}.


Finite-size physics is essential in this context. 
In particular, curvature contributions (often scaling as $E_{\rm curv}\propto R\,\gamma$) introduce an additional energy penalty that can suppress the formation of very small clusters, thereby reshaping the allowed parameter space for finite-$A$ SQM droplets. 
Moreover, electrostatic screening and the characteristically low charge-to-mass ratio ($Z/A \approx 0.1$) expected for strangelet matter can strongly reduce the electron pressure support relative to ordinary nuclei, leading to a noticeably softer EOS at a given baryon density. 
To treat the associated microphysical uncertainties---including the bag constant ($B^{1/4}$), surface tension ($\sigma$), curvature coefficient ($\gamma$), and strange-quark mass---we adopt a Bayesian framework that enables systematic parameter estimation under explicit stability and nuclear-physics constraints \citep{Trotta2008, Steiner2010}.

Several previous studies have explored the macroscopic consequences of strangelet-
or quark-matter-based equations of state, providing important insights into the
possible phenomenology of exotic compact objects. 
\citet{Alford2012b} constructed \textit{strangelet dwarfs} using a Wigner--Seitz
equation of state combined with a generic parametrization for the strangelet energy,
obtaining equilibrium sequences with masses $\sim 10^{-5}$--$0.1\,M_\odot$ and radii of order $500$--$5000$~km. 
Building on this approach, \citet{Zapata2020} considered a crystalline lattice of
strangelets described by the Heiselberg (1993) mass formula—which includes a surface
term but neglects curvature contributions—and focused on tidal disruption radii and
gravitational-wave signatures of such \textit{strangelet crystal planets}. 
In a different direction, \citet{Kurban2022} searched for strange-quark-matter objects
among white dwarfs using a bulk strange-matter equation of state without finite-size
strangelet effects, identifying candidate configurations with radii of several
thousand kilometers. 
Similarly, \citet{Wang2021} discussed observational signatures of \textit{strange quark
planets} assuming a homogeneous bulk composition and simple scaling relations, while
\citet{DiClemente2024} reviewed the stability of hybrid strange dwarfs featuring a
macroscopic quark core surrounded by a normal-matter crust.

Although these studies establish the plausibility of strangelet- or quark-matter-rich
compact objects, several critical aspects remain unexplored. In particular, none of
the above works simultaneously incorporates:
(i)~an explicit curvature term ($\gamma$) in the strangelet energy, which becomes
especially important for light strangelets with baryon number $A \le 100$;
(ii)~a systematic Bayesian exploration of the MIT bag model parameters
($B^{1/4}$, $\sigma$, $\gamma$, $m_s$) under the absolute-stability requirement, enabling
a controlled propagation of microphysical uncertainties into the equation of state;
(iii)~rapid rotation up to the Keplerian mass-shedding limit, consistently coupled with a relativistic equation of state, which can
substantially inflate the observable equatorial radius; and
(iv)~a direct comparison between theoretical mass--radius sequences and the NASA
Exoplanet Archive, allowing the \textit{density gap} between atomic-matter planets and
nuclear-density strangelet configurations to be quantified.

In this context, we address all these points. By focusing on light strangelets
($A \le 100$) and consistently treating surface and curvature effects within a
Bayesian framework, we obtain robust predictions for a new class of ultra-compact
sub-stellar objects—\emph{Substellar Strangelet Objects (SSOs)}—that are distinct from
previously studied strangelet dwarfs or strange planets. We further demonstrate that
rapid relativistic rotation can extend the equatorial radii of SSOs toward the
observational domain of massive exoplanets, thereby identifying concrete and testable
targets for future microlensing and high-cadence photometric surveys. These combined limitations motivate the present study.

In this work we develop a self-consistent structural model for compact configurations built from light strangelets ($A \leq 100$). 
We combine a modified liquid-drop description of finite-size SQM with a Bayesian calibration of the underlying microphysical parameters, and construct the corresponding EOS using the Wigner--Seitz cell approximation for strangelets embedded in a degenerate electron background. 
We then solve the Tolman--Oppenheimer--Volkoff (TOV) equations to obtain equilibrium sequences and show that the reduced electron degeneracy pressure implied by low $Z/A$ drives the solutions toward extreme compactness. 
A key outcome is that, under the microphysical constraints adopted here, the resulting stable sequences are confined to the \emph{sub-stellar} regime, with characteristic masses of order $M \sim 0.04\,M_{\odot}$ and radii $R \sim 15$--$60~\mathrm{km}$, and therefore cannot account for solar-mass white dwarfs. 
Instead, the model predicts an ultra-compact sub-stellar branch that can be confronted with observations: we compare our $M$--$R$ sequences against confirmed exoplanets and brown dwarfs from the NASA Exoplanet Archive \citep{Akeson2013, NASAExoplanetArchive} and highlight a pronounced ``density gap'' separating atomic-matter objects from the nuclear-density configurations predicted here. 
Finally, we assess the impact of rapid relativistic rotation near the Keplerian mass-shedding limit, which can inflate the equatorial radius and extend the accessible $M$--$R$ domain, providing an upper-envelope set of targets for future microlensing and high-cadence photometric searches.

This paper is organized as follows: 
In Section~2 we present the finite-size strangelet model and the Bayesian calibration strategy adopted to delineate the viable microphysical parameter space. 
In Section~3 we construct the EOS within the Wigner--Seitz framework and solve the TOV equations to obtain static equilibrium sequences. 
In Section~4 we quantify the effect of rapid relativistic rotation on the predicted mass--radius relation. 
We conclude in Section~5 with a discussion of observational implications, limitations of the present treatment, and prospects for detecting ultra-compact sub-stellar SQM candidates.

\vspace*{5mm}
\section{Microscopic Model}
\subsection{Properties of small strangelets ($A \le 100$)} \label{sec:strangelets}
To model the energetics of finite-sized strange quark matter (SQM), we adopt an extended MIT bag model within a liquid-drop approximation \citep{Farhi1984}. The total energy of a strangelet with baryon number $A$ and radius $R$ is written as a sum of bulk, finite-size, and electromagnetic contributions,
\begin{equation}
\label{eq:Etot}
\begin{aligned}
E_{\rm tot}(A,R) =\;& \frac{4}{3}\pi R^{3} B
+ E_{\rm kin}(A,R)
+ E_{\rm mass}(m_s) \\
&+ 4\pi R^{2}\sigma
+ 8\pi R\,\gamma
+ E_{\rm Coul}(A,R),
\end{aligned}
\end{equation}
where $B$ is the bag constant (vacuum pressure), $m_s$ is the strange-quark mass, and $\sigma$ and $\gamma$ denote the surface tension and curvature coefficient, respectively. 
In the literature, bag constants in the range $B^{1/4}\simeq 145$--$160~\mathrm{MeV}$ are typically compatible with bulk absolute stability of SQM under standard assumptions \citep{Chodos1974, Farhi1984, Madsen1999}. 
The kinetic term $E_{\rm kin}$ accounts for the relativistic Fermi motion of $u$, $d$, and $s$ quarks, while $E_{\rm mass}$ provides the leading correction associated with the finite strange-quark mass ($m_s \sim 90$--$150~\mathrm{MeV}$), which increases the energy relative to the massless approximation \citep{Alford2012}.
Finite-size geometrical effects are encoded in the surface and curvature terms. The surface contribution $4\pi R^{2}\sigma$ penalizes the SQM--vacuum interface, with typical theoretical estimates $\sigma \sim 10$--$50~\mathrm{MeV/fm^{2}}$ \citep{Madsen1993, Alford2012}. The curvature term $8\pi R\,\gamma$ becomes particularly important at low baryon number, where strong curvature can dominate the finite-size energy budget; representative ranges are $\gamma \sim 5$--$20~\mathrm{MeV/fm}$ \citep{Madsen1993}. Finally, the Coulomb self-energy is modeled as
\begin{equation}
E_{\rm Coul}(A,R) = \frac{3}{5}\,\frac{Z^{2}\alpha\,\hbar c}{R},
\end{equation}
where $Z$ is the strangelet charge and screening effects may cause $Z(A)$ to deviate from the bulk limit \citep{Heiselberg1993}.

Equation~(\ref{eq:Etot}) provides a compact parametrization that isolates the physical origin of each contribution and makes explicit the parameters that will be calibrated in the Bayesian analysis below. In particular, at $A\lesssim 100$, surface and curvature terms can raise $E/A$ significantly relative to the bulk value, so finite-$A$ properties must be treated consistently with the adopted stability criteria.

\subsection{Bayesian calibration of microphysical parameters}
\label{sec:bayesian}
The microphysical parameters governing finite-size strange quark matter are subject to significant theoretical uncertainties, particularly in the low-baryon-number regime where surface and curvature effects become relevant. Rather than adopting fixed values, we employ a Bayesian framework to systematically explore the allowed parameter space and to quantify the associated uncertainties. In this context, Bayesian inference is used as a \emph{calibration and consistency tool}, enabling a controlled propagation of microphysical uncertainties into the macroscopic stellar models.

The parameter space explored includes the bag constant ($B^{1/4}$), surface tension ($\sigma$), curvature coefficient ($\gamma$), and strange-quark mass ($m_s$). The analysis is constrained by standard nuclear-physics consistency requirements, most notably the absolute stability condition for bulk strange quark matter ($E/A < 930$ MeV) and the requirement that ordinary nuclei remain stable against decay into two-flavor quark matter. These conditions define a physically motivated stability window within which viable parameter combinations must reside \citep{Farhi1984,Steiner2010,Steiner2013}.

Within this framework, the Bayesian approach serves two complementary purposes. First, it identifies regions of parameter space compatible with microphysical stability, avoiding ad hoc parameter choices. Second, it provides a quantitative estimate of the dispersion and correlations among admissible parameters, which are then propagated into the equation of state and the macroscopic equilibrium sequences. Importantly, this procedure is not intended as a direct statistical fit to the full population of observed compact objects, but rather as a means of consistently incorporating microphysical uncertainties into the modeling of strangelet-rich matter \citep{Trotta2008}.

The representative parameter values adopted in this work correspond to central tendencies of the resulting posterior distributions and are used throughout the subsequent sections to construct the equation of state and compute static and rotating equilibrium configurations. The qualitative conclusions of the present study—namely, the confinement of stable configurations to the sub-stellar mass regime and the emergence of ultra-compact planetary-mass objects—are robust against moderate variations within the calibrated parameter space, as demonstrated explicitly in Sections \ref{sec:macroscopic} and \ref{sec:rotation}.

The Bayesian exploration yields the following representative parameter values,
which are adopted as reference throughout this work:
\begin{equation}
\label{eq:bayes_params}
\begin{aligned}
B^{1/4} &= 157.27 \pm 4.73~\mathrm{MeV}, \\
\sigma  &= 24.21 \pm 12.67~\mathrm{MeV/fm^{2}}, \\
\gamma  &= 24.31 \pm 9.88~\mathrm{MeV/fm}, \\
m_s     &= 131.46 \pm 28.89~\mathrm{MeV}.
\end{aligned}
\end{equation}
\subsection{Representative strangelet properties in vacuum}
\label{sec:props}
Using the median calibrated parameters of Eq.~(\ref{eq:bayes_params}), we compute illustrative vacuum properties for selected strangelet baryon numbers. Table~\ref{tab:strangelet_props} summarizes the resulting radii, charges, and energies per baryon.

As seen in Table~\ref{tab:strangelet_props}, finite-size strangelets exhibit a characteristically low charge-to-mass ratio, $Z/A \simeq 0.1$, markedly smaller than that of ordinary nuclei ($Z/A \sim 0.5$). 
This ``charge deficit'' reduces the electron density required for charge neutrality in a strangelet lattice and therefore lowers the electron degeneracy pressure for a given baryon density. 
We emphasize that the finite-size corrections included in Eq.~(\ref{eq:Etot}) can raise $E/A$ above the bulk stability threshold at small $A$; in our framework the bulk stability criteria are used to delimit the calibrated microphysical parameter space, while the finite-$A$ properties determine the EOS relevant for macroscopic equilibrium.
\begin{table}[t]
\centering
\caption{Illustrative vacuum properties of finite-$A$ strangelets evaluated with the median calibrated parameters.}
\label{tab:strangelet_props}
\begin{tabular}{lcccc}
\hline\hline
$A$ & $R_{\rm str}$ (fm) & $Z$ & $Z/A$ & $E/A$ (MeV) \\
\hline
12  & 1.86 & 1.2  & 0.100 & 1191.20 \\
36  & 2.79 & 3.6  & 0.100 & 1113.06 \\
56  & 3.26 & 5.6  & 0.100 & 1091.86 \\
100 & 4.00 & 10.0 & 0.100 & 1070.15 \\
\hline
\end{tabular}
\tablecomments{Values computed with $B^{1/4} = 157.27~\mathrm{MeV}$, $\sigma = 24.21~\mathrm{MeV/fm^{2}}$, $\gamma = 24.31~\mathrm{MeV/fm}$, and $m_s=131.46~\mathrm{MeV}$.}
\end{table}


\subsection{Equation of state: Wigner--Seitz strangelet matter}
\label{sec:eos}
To describe the thermodynamics of a strangelet--electron plasma, we adopt the Wigner--Seitz (WS) cell approximation \citep{Glendenning2000,Alford2012}. The system is partitioned into identical spherical cells of radius $R_{\rm cell}$, each containing a single strangelet at the center and a uniform, degenerate electron background. Global charge neutrality in each cell implies
\begin{equation}
n_e = \frac{Z}{V_{\rm cell}}, \qquad
V_{\rm cell} = \frac{4\pi}{3}R_{\rm cell}^{3},
\end{equation}
and the baryon density is
\begin{equation}
n_b = \frac{A}{V_{\rm cell}}.
\end{equation}
The total energy density is then written as
\begin{equation}
\label{eq:epsilon_ws}
\epsilon(n_b) =
\frac{E_{\rm str}(A,R_{\rm str}) + E_{\rm lat}(Z,R_{\rm cell})}{V_{\rm cell}}
+ \epsilon_e(n_e),
\end{equation}
where $E_{\rm str}$ is the strangelet rest energy including surface and curvature terms, $E_{\rm lat}$ denotes the lattice (Coulomb) correction, and $\epsilon_e$ is the energy density of a degenerate electron gas. The electron pressure $P_e$ and $\epsilon_e(n_e)$ follow from standard Fermi-gas expressions \citep{Chandrasekhar1939, Salpeter1961}. In practice, the total pressure is dominated by the electrons over the density range of interest, with subleading lattice contributions (included through $E_{\rm lat}$) \citep{Alford2012}.
\begin{figure}[t]
    \centering
    \includegraphics[width=\columnwidth]{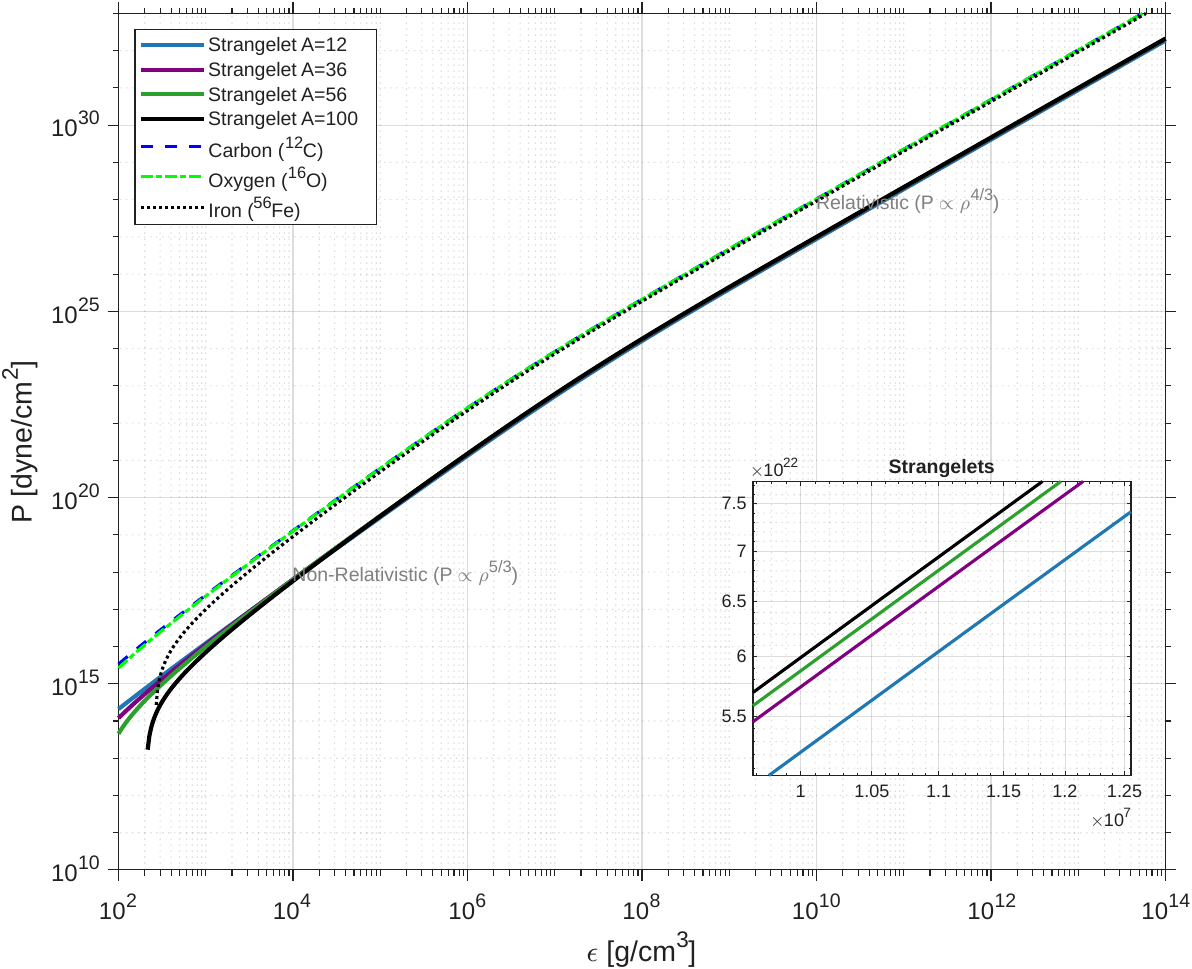}
    \caption{Equation of state (EOS) comparison between strangelet matter (representative $A=12, 36, 56, 100$) and standard baryonic matter ($^{12}$C, $^{16}$O, $^{56}$Fe).}
    \label{fig:eos_summary}
\end{figure}

Figure~\ref{fig:eos_summary} highlights two key physical features of strangelet-rich matter. 
First, for a given mass--energy density, the equation of state of strangelet matter is systematically \emph{softer} than that of ordinary hadronic matter (e.g., carbon, oxygen, or iron). 
This behavior originates from the anomalously low charge-to-baryon ratio of strangelets ($Z/A \ll 0.5$), which strongly suppresses the number density of degenerate electrons required for charge neutrality. 
Because the pressure support in the Wigner--Seitz picture is dominated by the electron degeneracy pressure \citep{ShapiroTeukolsky1983}, a reduced electron fraction directly translates into a lower pressure at fixed energy density compared to standard baryonic matter. 
As a result, ordinary hadronic matter can sustain substantially higher pressures than strangelet matter at the same density.

Second, the inset of Fig.~\ref{fig:eos_summary} (strangelets) shows that the stiffness of the strangelet equation of state depends sensitively on the strangelet baryon number $A$. 
For a fixed density, increasing $A$ leads to a systematically higher pressure, reflecting the reduced relative importance of surface and curvature energy contributions for larger strangelets. 
Consequently, matter composed of larger strangelets (e.g., $A=100$) provides more pressure support against gravity than matter composed of smaller strangelets (e.g., $A=12$). 
This trend implies that compact objects built from larger-$A$ strangelets are able to support larger maximum masses and radii, a feature that will become evident in the macroscopic mass--radius sequences discussed in the following section.
\vspace*{5mm}
\section{Macroscopic Structure of Sub-stellar Strangelet Objects}
\label{sec:macroscopic}

With the microscopic description of strangelets and the corresponding EOS (Section~\ref{sec:eos}) in hand, we now determine the macroscopic properties of self-gravitating equilibrium configurations. We restrict ourselves in this section to \emph{static}, spherically symmetric models in general relativity. For convenience we work in geometrized units ($G=c=1$) throughout; all masses and radii are converted to $M_\odot$ and km for presentation.
We adopt the standard static, spherically symmetric metric
\begin{equation}
ds^{2}=-e^{2\Phi(r)}dt^{2}+e^{2\Lambda(r)}dr^{2}+r^{2}d\Omega^{2},
\end{equation}
where $\Phi(r)$ and $\Lambda(r)$ are metric potentials and $d\Omega^{2}=d\theta^{2}+\sin^{2}\theta\,d\phi^{2}$ is the solid angle element. Solving Einstein's field equations for a perfect fluid yields the Tolman--Oppenheimer--Volkoff (TOV) equations \citep{tolman1939static, oppenheimer1939massive}:
\begin{equation}
\label{eq:tov_dp}
\frac{dP}{dr}=-\frac{(\epsilon+P)\left(m+4\pi r^{3}P\right)}{r\left(r-2m\right)},
\end{equation}
together with the mass continuity equation
\begin{equation}
\label{eq:tov_dm}
\frac{dm}{dr}=4\pi r^{2}\epsilon,
\end{equation}
closed by an EOS $P=P(\epsilon)$ (Section~\ref{sec:eos}). For a chosen central energy density $\epsilon_c$, we integrate Eqs.~(\ref{eq:tov_dp})--(\ref{eq:tov_dm}) from $r=0$ outward with $m(0)=0$ and $P(0)=P_c$, stopping at the surface $r=R$ defined by $P(R)=0$. The gravitational mass is then $M\equiv m(R)$.

\begin{figure}[t]
    \centering
    \includegraphics[width=\columnwidth]{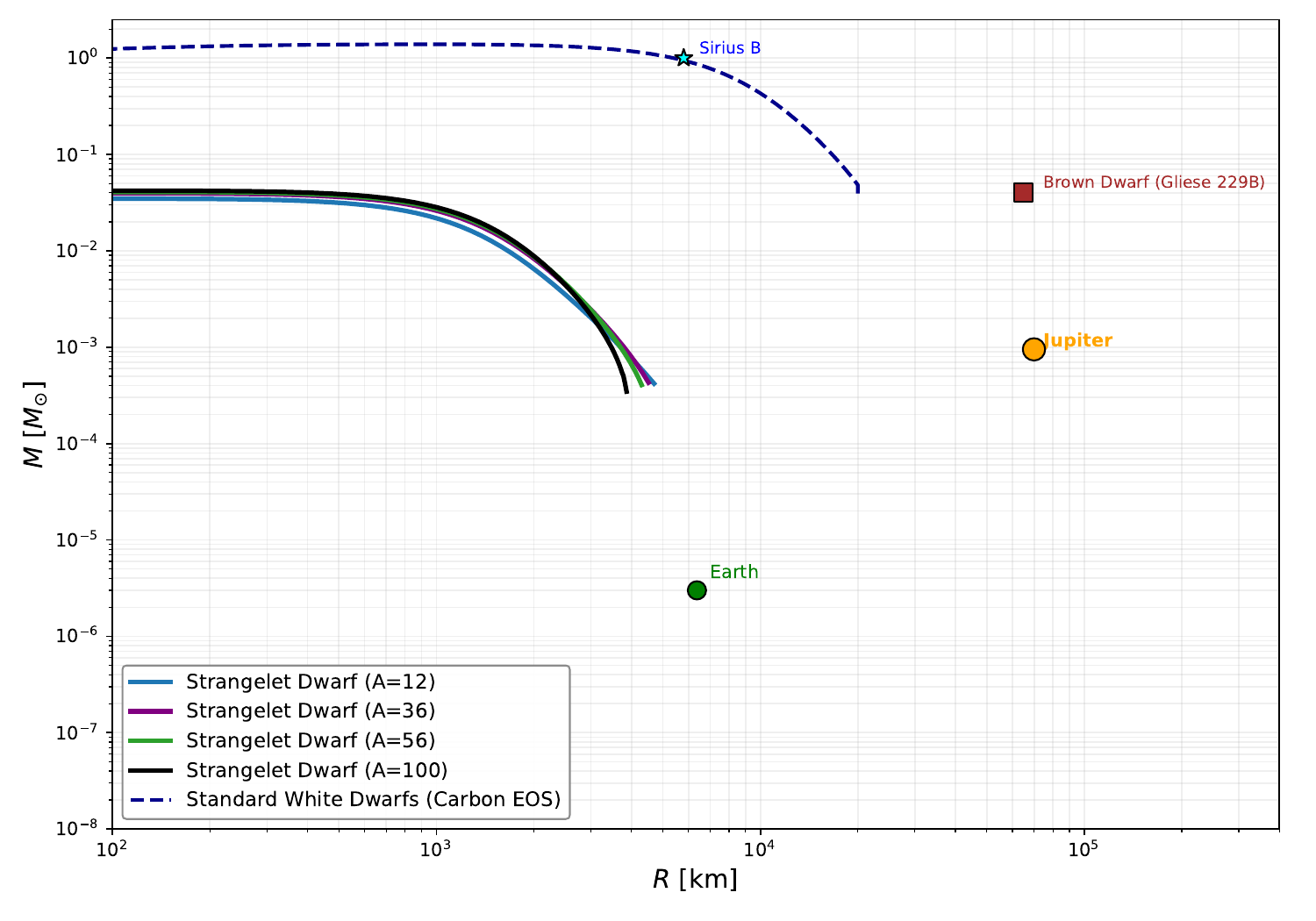}
    \caption{
    Static mass--radius relations for sub-stellar strangelet-supported objects (representative $A=12,36,56,100$ sequences) compared with the standard carbon--oxygen white dwarf sequence (dashed blue line). Reference bodies (Earth, Jupiter, and the brown dwarf Gliese~229B) are shown for scale. The figure emphasizes the orders-of-magnitude separation in radius between atomic-matter objects and the ultra-compact strangelet branch, defining an observational ``density gap'' in the $M$--$R$ plane.
    }
    \label{fig:mr_sequence}
\end{figure}

Our primary interest is the mass--radius ($M$--$R$) relation of strangelet-supported configurations. Figure~\ref{fig:mr_sequence} shows the static sequences obtained from the strangelet EOS (representative $A$ values) and compares them with the standard carbon white dwarf sequence, together with reference bodies (Earth, Jupiter, and a typical brown dwarf).
The contrast with ordinary white dwarfs is immediate: while Carbon white dwarfs populate radii of thousands of kilometers and reach the Chandrasekhar mass scale ($\sim 1.4\,M_\odot$), the strangelet-rich branch is confined to a \emph{sub-stellar} window with characteristic radii of order $10^{3}$--$10^{4}$ km.
In our models, the stable sequences typically attain maximum masses in the range $M_{\max}\simeq 0.02$--$0.08\,M_\odot$ with characteristic radii of order $10^{3}$--$10^{4}$ km, remaining orders of magnitude smaller than those of ordinary planets at comparable masses.

This extreme compactness follows directly from the microphysics of strangelet matter. Ordinary baryonic compositions (e.g., C/O or Fe) have $Z/A\simeq 0.5$, which implies a relatively large electron number density at charge neutrality and therefore substantial electron degeneracy pressure. By contrast, strangelets have a pronounced ``charge deficit'' ($Z/A\sim 0.1$; Section~\ref{sec:props}), so achieving neutrality requires far fewer electrons per baryon. Because the pressure support in the Wigner--Seitz picture is dominated by the degenerate electron gas, a reduced electron fraction softens the EOS and forces the configurations to compress to much higher densities before the degeneracy pressure can balance gravity. As a result, our solutions occupy a region of the $M$--$R$ plane characterized by planet-like masses but \emph{nuclear-scale compactness}, thereby defining a pronounced observational ``density gap'' between atomic-matter objects (planets/brown dwarfs/white dwarfs) and the ultra-compact strangelet branch.

To further contextualize the internal structure of these theoretical objects, it is instructive to examine their macroscopic central densities $\rho_c$. Unlike ordinary rocky or gaseous planets, which maintain central densities of order $\mathcal{O}(10^{1})~\mathrm{g\,cm^{-3}}$, the strange sub-stellar objects (SSOs) spaning from planetary masses ($\sim10^{-6}$--$10^{-2}\,M_\odot$) up to the sub-stellar regime ($\lesssim10^{-1}\,M_\odot$) exhibit extreme central densities spanning $\epsilon_c \sim 10^{6}$--$10^{11}~\mathrm{g\,cm^{-3}}$. This highly compact macroscopic state is a direct consequence of the low charge fraction of strangelet matter: with $Z/A\ll 1$, gravity must compress the configuration to large baryon densities in order to pack enough electrons into a given volume to supply the degeneracy pressure required for hydrostatic support.
It is crucial to distinguish this \emph{macroscopic} central density from the \emph{microscopic} density inside an individual strangelet. While the volume-averaged core density lies in the $10^{6}$--$10^{11}~\mathrm{g\,cm^{-3}}$ regime, the local density \emph{inside} a strangelet (within the MIT bag) remains nearly incompressible, close to an extended nuclear saturation density, $\sim 4\times 10^{14}~\mathrm{g\,cm^{-3}}$. The macroscopic star can therefore be understood as a lattice of ultra-dense strangelet droplets embedded in a highly compressed, neutralizing degenerate electron gas, which explains why radii remain in the $10^{3}$--$10^{4}$ km range even for planet-like masses.
\begin{figure}[t]
    \centering
    \includegraphics[width=\columnwidth]{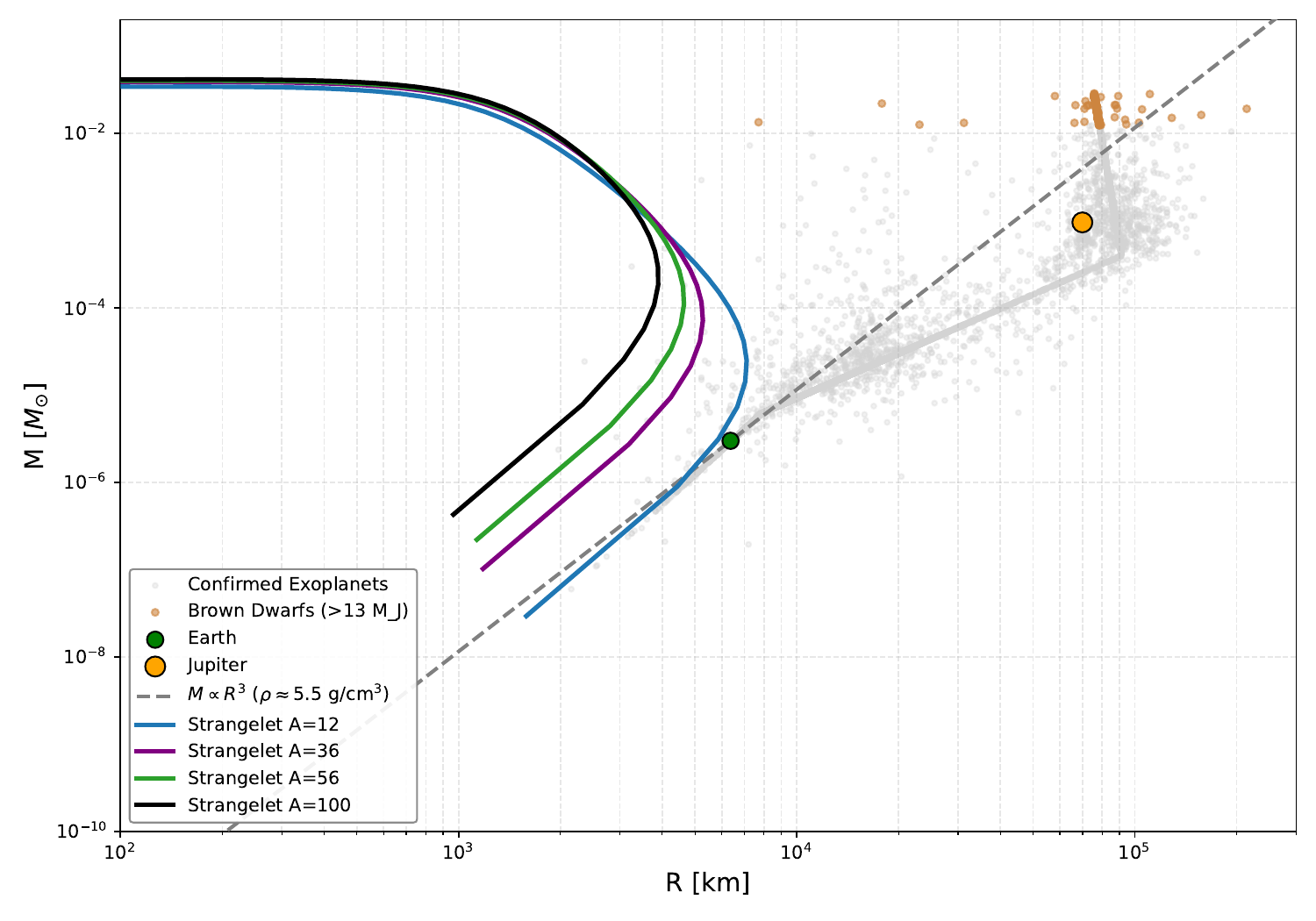}
    \caption{
    Mass--radius comparison between strangelet-rich sub-stellar models and observational catalogs. Solid colored curves show static theoretical sequences for representative strangelet matter models. Gray circles denote confirmed exoplanets, while brown markers indicate brown dwarfs ($M>13\,M_J$) from the NASA Exoplanet Archive \citep{NASAExoplanetArchive, Akeson2013}. Earth and Jupiter are shown as reference points. The thin reference line corresponds to an Earth-density constant-density scaling ($M=\frac{4\pi}{3}\rho_\oplus R^3$), illustrating the approximate $M\propto R^3$ trend followed by ordinary planets at low densities. The strangelet branch lies orders of magnitude below this locus, reflecting its ultra-compact (nuclear-density) nature.
    }
    \label{fig:observations}
\end{figure}

Figure~\ref{fig:observations} compares our theoretical sequences with confirmed exoplanets and brown dwarfs from the NASA Exoplanet Archive \citep{NASAExoplanetArchive, Akeson2013}. In the observational domain, low- and intermediate-mass planets broadly track approximately constant-density scalings, $M\propto R^{3}$, because their bulk densities vary within a limited range set by composition (rock/ice/gas) and modest self-compression. For guidance, the figure includes a reference line corresponding to an Earth-like mean density, $\bar{\rho}\approx \rho_\oplus$, i.e.
\begin{equation}
\label{eq:rhoearth_line}
M = \frac{4\pi}{3}\,\rho_\oplus\,R^{3},
\end{equation}
which provides a convenient baseline for interpreting the exoplanet cloud at low densities.

In sharp contrast, strangelet-rich configurations occupy radii of only $R\simeq 10^{2}$--$\sim10^{4}$ km over a broad range of planet-like masses. In other words, objects with masses comparable to giant planets or low-mass brown dwarfs would sit many orders of magnitude below the planetary constant-density locus (Eq.~\ref{eq:rhoearth_line}). This behavior reflects the fundamentally different pressure support mechanism: because $Z/A$ is anomalously small, the degenerate electron pressure is suppressed at a given baryon density, so hydrostatic equilibrium is reached only after the configuration is compressed to much larger densities (Section~\ref{sec:eos}). At the lowest masses, where self-gravity is weakest, the strangelet branch can locally resemble a ``constant-density'' scaling in log--log space (i.e., $M\propto R^3$), but with an \emph{enormously larger} effective density scale than planets, producing a curve that is roughly parallel to planetary constant-density lines while being shifted to radii smaller by orders of magnitude.

The observational implication is that SSOs would be essentially invisible to standard transit depth surveys because of their tiny cross-sections, yet they could be detected through gravitational microlensing \cite{Griest2011}, pulsar/exoplanet timing, or other high-cadence photometric signatures where they act as point-like massive lenses. These diagnostics naturally follow from their planet-like masses combined with $10^{3}$--$10^{4}$ km radii and extreme central densities discussed above.

\section{Rotation and the Keplerian Limit}
\label{sec:rotation}
To assess how rapid rotation modifies the observable properties of strangelet-rich sub-stellar objects, we consider stationary and axisymmetric configurations. While our static spherically symmetric models (Section 3) are obtained by integrating the full relativistic Tolman-Oppenheimer-Volkoff (TOV) equations, it is crucial to note the characteristic compactness of these sub-stellar objects. For our relevant mass range ($M \lesssim 0.1 M_{\odot}$) and radii of order $10^3 - 10^4$ km, the dimensionless compactness parameter is exceedingly small ($GM/Rc^2 \sim 10^{-5}$). 

Because spacetime curvature is strictly negligible in this macroscopic regime, computing rotating equilibria in full General Relativity (e.g., including frame-dragging metrics) is unnecessary. Instead, we can safely and accurately model these rapidly rotating configurations using the classical Self-Consistent Field (SCF) method developed by \citet{Hachisu1986}. To ensure thermodynamic consistency with our static TOV models, we incorporate the fully relativistic equation of state into the SCF framework through the relativistic enthalpy integral.

This pseudo-Newtonian approach not only ensures robust convergence at extreme rotation rates but also clarifies the physical interpretation of the stellar boundary. Because the spatial metric in our numerical grid remains flat (Euclidean), there is no geometric spatial contraction. Consequently, the coordinate equatorial radius ($R_{eq}$) extracted dynamically as the eigenvalue in the SCF iteration is mathematically identical to the physical circumferential radius ($R_{\text{circ}} = \sqrt{g_{\phi\phi}} = R_{eq}$). The radii reported in our mass-radius diagrams thus represent the true, observable cross-sectional dimensions of the configurations, directly comparable with astrometric and transit observations.

Uniformly rotating configurations admit a maximum angular frequency set by the mass-shedding (Keplerian) limit. Physically, this limit is reached when a fluid element at the equator becomes marginally bound: the stellar angular velocity equals the orbital frequency of a test particle at the equatorial surface. Any further increase in the rotation rate would lead to mass loss from the equator. The corresponding Keplerian angular frequency $\Omega_{K}$ therefore provides a global constraint that depends on both the EOS and the centrifugal deformation of the star \citep{Hachisu1986, Friedman2013}. 

A useful order-of-magnitude estimate is given by the expression:
\begin{equation}
f_{K} \simeq \frac{\Omega_{K}}{2\pi} \approx \frac{1}{2\pi}\sqrt{\frac{GM}{R_{eq}^3}},\label{eq:fK_newton}
\end{equation}

Because strangelet-rich objects have kilometre-scale radii at planet-like or sub-stellar masses (Section~\ref{sec:macroscopic}), Eq.~(\ref{eq:fK_newton}) implies that the Keplerian frequencies can be very large compared to ordinary planets: typical periods can fall in the range of seconds down to fractions of a second depending on $(M,R_{\rm eq})$. This ``hyper-compactness'' is a direct consequence of the microphysics: the low charge-to-mass ratio suppresses the electron pressure support, forcing the equilibrium to occur at much higher densities (Section~\ref{sec:eos}), which deepens the gravitational potential well and allows the object to withstand extreme centrifugal support before shedding mass.

As the Keplerian limit is approached, configurations develop a pronounced oblateness and exhibit a substantial increase in equatorial radius relative to the non-rotating case. 
This behavior arises from centrifugal support and the global redistribution of matter toward the equatorial region, 
and is typically accompanied by a modest increase in the maximum gravitational mass supported by the configuration \citep{Komatsu1989, Cook1994, FriedmanStergioulas2013}. In the context of strangelet-rich sub-stellar objects, these rotational effects extend the accessible mass--radius ($M$--$R$) domain without violating relativistic hydrodynamic stability. Rapidly rotating models near $\Omega_{\rm K}$ should therefore be interpreted as \emph{extremal} solutions that provide an upper envelope in the $M$--$R$ plane rather than representing generic equilibrium states.

\begin{figure}[t]
    \centering
    \includegraphics[width=\columnwidth]{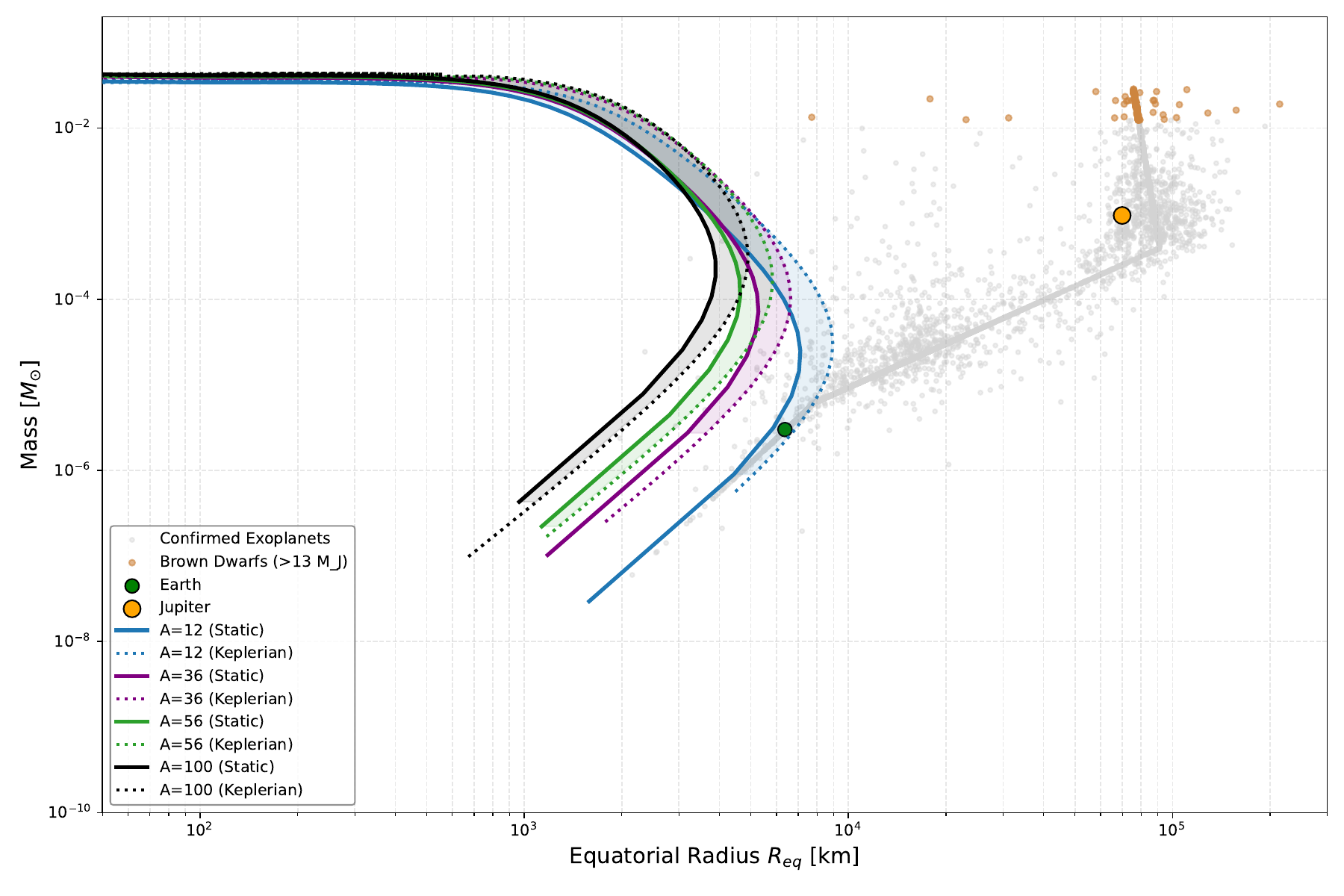}
    \caption{
    Mass--radius comparison between strangelet-rich sub-stellar models and observational data. Solid curves show non-rotating (spherical) equilibrium sequences, while dashed curves correspond to rapidly rotating configurations near the Keplerian (mass-shedding) limit, plotted using the equatorial radius $R_{\rm eq}$, for representative effective baryon numbers $A$. Gray circles denote confirmed exoplanets and brown markers indicate brown dwarfs ($M>13\,M_J$) from the NASA Exoplanet Archive. Earth and Jupiter are included as reference points. Rotation produces a pronounced inflation of $R_{\rm eq}$ and extends the theoretical sequences toward the observational domain of massive exoplanets, providing an upper envelope in the $M$--$R$ plane.}
    \label{fig:MR_rotation}
\end{figure}

Figure~\ref{fig:MR_rotation} shows the mass--radius relations for strangelet-rich objects with representative effective baryon numbers $A$, comparing static configurations (solid curves) with rapidly rotating models near the Keplerian limit (dashed curves). For observational context, confirmed exoplanets and brown dwarfs from the NASA Exoplanet Archive are also shown. The main qualitative result is that rotation can inflate $R_{\rm eq}$ by a significant factor, shifting the theoretical curves toward larger radii and bringing the most extended configurations closer to the region populated by massive exoplanets \citep{Seager2007}.

In the absence of rotation, the strangelet sequences occupy an ultra-compact region of the mass--radius plane, with small radii across a broad mass range. This reflects the high mean densities imposed by the strangelet EOS and the reduced electron pressure support associated with low $Z/A$ (Sections~\ref{sec:props} and \ref{sec:eos}). Consequently, static strangelet-rich objects are systematically displaced to radii many orders of magnitude smaller than those of observed exoplanets at comparable mass.

Including rapid rotation substantially modifies this picture. Near $\Omega_{\rm K}$, centrifugal support and oblateness redistribute matter toward the equator, increasing $R_{\rm eq}$ and slightly raising the maximum supported mass \citep{Cook1994, FriedmanStergioulas2013}. In the $M$--$R$ diagram, these effects appear as a systematic shift of the theoretical curves toward larger radii. For certain $A$ values and masses, the Keplerian sequences partially overlap with the locus of massive exoplanets, indicating that strangelet-rich objects could, in principle, attain radii comparable to some observed bodies \emph{only} if they rotate sufficiently close to the mass-shedding threshold.

A useful consistency check is to compare the rotation rates required for substantial radius inflation with the inferred spin periods of observed exoplanets. Giant planets and compact exoplanets are generally expected to rotate with periods from several hours to a few days, depending on formation history and tidal coupling to the host star \citep{Snellen2014, Showman2017}. By contrast, the near-Keplerian configurations that maximize $R_{\rm eq}$ in our models correspond to much higher spin frequencies because of the kilometre-scale radii of SSOs. 
Using Eq.~(\ref{eq:fK_newton}) as an order-of-magnitude guide, which is highly reliable given the low compactness of these systems, 
the Keplerian periods for our sequences can plausibly range from a few seconds down to fractions of a second across the relevant $(M,R_{\rm eq})$ domain. This implies that only a subset of systems---for instance, young objects or those with weak tidal synchronization---could plausibly approach the extremal Keplerian envelope.

Overall, Figure~\ref{fig:MR_rotation} delineates the region of parameter space where strangelet-rich sub-stellar objects remain observationally viable when rotation is included. The key message is that rotation can move the theoretical branch toward the exoplanet domain, but only at extremal spin rates near the relativistic mass-shedding limit; thus, rapidly rotating models should be interpreted as upper bounds on radii for a given mass and EOS rather than typical configurations.
\vspace*{5mm}
\section{Discussion and Conclusions}
This study provides a reassessment of the role of light strangelets ($A \leq 100$) in compact sub-stellar and planetary-scale structures. Motivated by the reported mass--radius outliers in the \textit{Gaia} DR3 era \citep{GaiaDR3_2023,Althaus2022}, we find a robust theoretical constraint: once finite-size effects (surface and curvature terms) are consistently included, equilibrium configurations built from light strangelet matter are confined to the \emph{sub-stellar} regime ($M \lesssim 0.08\,M_{\odot}$). Consequently, light strangelets cannot account for solar-mass white dwarfs, irrespective of the macroscopic treatment adopted.

A methodological contribution of this work is the use of a Bayesian framework to explore and constrain the microphysical parameter space controlling the stability window of strange matter at low baryon number. In contrast to purely deterministic parameter choices \citep{Glendenning1995b}, the Bayesian analysis allows us to quantify parameter uncertainties under the standard absolute-stability requirement ($E/A < 930$ MeV) and the requirement that ordinary nuclei remain stable against decay into two-flavor quark matter \citep{Trotta2008,Steiner2010}. Within this constrained space, the intrinsically low charge-to-mass ratio of strangelets ($Z/A \sim 0.1$) emerges as the key driver of the macroscopic outcome: reduced electron fractions suppress electron degeneracy pressure, shifting the maximum supported masses far below the Chandrasekhar scale for carbon--oxygen matter.

The main physical implication is the appearance of a distinct family of ultra-compact objects at planet-like masses. In the mass--radius plane, ordinary exoplanets and brown dwarfs occupy radii of order $10^{4}$--$10^{5}$ km, while the strangelet-rich sequences remain confined to radii of $\simeq 10^{2}$--$\sim10^{4}$ km over comparable masses. This separation defines a pronounced \emph{density gap} between atomic-matter objects and the nuclear-compact branch predicted here. We refer to these configurations as \emph{Sub-stellar Strangelet Objects} (SSOs).

Potential formation channels for SSOs include: fragmentation of strange-star crust material in core-collapse events \citep{Madsen1999}, tidal ejection in strange-star mergers \citep{Bauswein2009}, and primordial production associated with the QCD phase transition in the early Universe \citep{Witten1984,Lynn1990}. Although the detailed efficiencies of these channels remain uncertain, each provides a plausible route to populating the sub-stellar mass range with compact strangelet-rich remnants.

Including rapid rotation while consistently accounting for relativistic thermodynamics substantially broadens the accessible mass--radius domain. Models approaching the Keplerian mass-shedding limit exhibit significant equatorial-radius inflation and a modest increase in the maximum gravitational mass \citep{Cook1994,FriedmanStergioulas2013}. While rotation does not erase the density gap, it shifts the \emph{upper envelope} of the strangelet branch toward larger radii and can yield partial overlap with the observational parameter space of massive exoplanets. Importantly, these near-Keplerian configurations are \emph{extremal} rotational states and should be interpreted as theoretical upper bounds, rather than representative equilibrium solutions. Nevertheless, they demonstrate that rotation is essential for a realistic assessment of the observational viability of strangelet planetary objects.

From an observational standpoint, SSOs combine planet-like masses with kilometre-scale radii and extreme compactness. Their tiny geometric cross-sections imply that they would produce extremely shallow transits and could evade conventional transit surveys. Instead, their most promising signatures are expected in channels sensitive to mass rather than area, such as gravitational microlensing, high-cadence photometric monitoring (including transit timing anomalies), and precision timing applications.

Two additional physical effects may help preserve the ultra-compact nature of SSOs over long timescales. First, strong-interaction binding makes strange quark matter self-bound, so these objects can in principle withstand much shorter spin periods than ordinary rocky planets before reaching mass shedding \citep{Weber2005}. Second, a strong Coulomb barrier at a bare quark surface can suppress accretion of ordinary baryonic material from the interstellar medium, limiting the growth of an extended atomic crust and helping maintain an ultra-compact configuration \citep{Alcock1986,Usov1997}. These features distinguish SSOs from conventional planets and brown dwarfs and motivate dedicated searches optimized for compact, massive lenses.

Several limitations should be kept in mind. The liquid-drop treatment of light strangelets neglects shell effects, additional finite-size corrections beyond the leading curvature term, and possible color-superconducting phases that could modify the stability window and EOS at higher densities or larger baryon numbers \citep{Alford2012}. Moreover, a fully consistent characterization of rotating equilibria---including moments of inertia and multipole moments---would benefit from complementary treatments (e.g., slow-rotation expansions) and systematic comparisons across numerical rotation solvers \citep{HartleThorne1968,Cook1994}. Finally, formation scenarios and population synthesis for SSOs remain open problems that will require connecting microphysical production yields to astrophysical environments and dynamical capture/ejection processes.

In summary, although light strangelets do not resolve the puzzle of overmassive white dwarfs, they robustly predict a previously unexplored landscape of ultra-compact sub-stellar matter. The resulting SSOs constitute a distinct population separated from atomic-matter planets and brown dwarfs by a large density gap, and rotation can expand their accessible radii toward the massive-exoplanet domain at extremal spin rates. These findings challenge standard mass--radius paradigms and motivate observational strategies aimed at detecting compact, rapidly rotating, planet-mass objects beyond the conventional framework of atomic matter. Future high-precision surveys will be essential to test this scenario.

\vspace*{5mm}
\section*{Acknowledgements}
The authors would like to express their gratitude to the Universidad Tecnológica del Perú (UTP) for the institutional support and access to the computing resources necessary for the development of this work. 
We also thank the NASA Exoplanet Archive for providing the high-quality observational data that made our comparative analysis possible. This research was conducted independently and did not receive any specific grant from funding agencies in the public, commercial, or not-for-profit sectors.

\bibliography{biblio}

@ARTICLE{Chandrasekhar1931,
       author = {{Chandrasekhar}, S.},
        title = "{The Maximum Mass of Ideal White Dwarfs}",
      journal = {\apj},
         year = 1931,
       volume = {74},
        pages = {81},
          doi = {10.1086/143324}
}

@ARTICLE{GaiaDR3_2023,
       author = {{Gaia Collaboration} and {Vallenari}, A. and others},
        title = "{Gaia Data Release 3. Summary of the content and survey properties}",
      journal = {\aap},
         year = 2023,
       volume = {674},
          eid = {A1},
        pages = {A1},
          doi = {10.1051/0004-6361/202243940}
}

@ARTICLE{Bodmer1971,
       author = {{Bodmer}, A.~R.},
        title = "{Collapsed Nuclei}",
      journal = {\prd},
         year = 1971,
       volume = {4},
        pages = {1601},
          doi = {10.1103/PhysRevD.4.1601}
}

@ARTICLE{Witten1984,
       author = {{Witten}, Edward},
        title = "{Cosmic separation of phases}",
      journal = {\prd},
         year = 1984,
       volume = {30},
        pages = {272},
          doi = {10.1103/PhysRevD.30.272}
}

@article{Trotta2008,
  title={Bayes in the sky: Bayesian inference and model selection in cosmology},
  author={Trotta, Roberto},
  journal={Contemporary Physics},
  volume={49},
  number={2},
  pages={71--104},
  year={2008},
  publisher={Taylor \& Francis}
}

@article{Steiner2010,
  title={The equation of state from observed masses and radii of neutron stars},
  author={Steiner, Andrew W and Lattimer, James M and Brown, Edward F},
  journal={The Astrophysical Journal},
  volume={722},
  number={1},
  pages={33},
  year={2010},
  publisher={IOP Publishing}
}

@ARTICLE{Glendenning1995a,
       author = {{Glendenning}, N.~K. and {Kettner}, Ch. and {Weber}, F.},
        title = "{From stars to strange stars}",
      journal = {\apj},
         year = 1995,
       volume = {450},
        pages = {253},
          doi = {10.1086/176137}
}

@ARTICLE{Glendenning1995b,
       author = {{Glendenning}, Norman K. and {Kettner}, Christian and {Weber}, Fridolin},
        title = "{Strange-matter stars}",
      journal = {Physical Review Letters},
         year = 1995,
       volume = {74},
        pages = {3519},
          doi = {10.1103/PhysRevLett.74.3519}
}

@ARTICLE{Madsen1993,
       author = {{Madsen}, Jes},
        title = "{Curvature contribution to the s-quark mass-dependent surface tension of strangelet drops}",
      journal = {\prd},
         year = 1993,
       volume = {47},
        pages = {5156},
          doi = {10.1103/PhysRevD.47.5156}
}

@ARTICLE{Alford2012,
       author = {{Alford}, Mark G. and {Han}, Sophia and {Reddy}, Sanjay},
        title = "{Strangelet fraction in the crust of strange stars}",
      journal = {Journal of Physics G},
         year = 2012,
       volume = {39},
          eid = {065201},
        pages = {065201},
          doi = {10.1088/0954-3899/39/6/065201}
}

@article{Alford2012b,
  author  = {Alford, Mark G. and Han, Sophia and Reddy, Sanjay},
  title   = {Strangelet dwarfs},
  journal = {Physical Review D},
  volume  = {85},
  number  = {4},
  pages   = {045003},
  year    = {2012},
  doi     = {10.1103/PhysRevD.85.045003}
}

@article{Zapata2020,
  author  = {Zapata, Joás and Negreiros, Rodrigo},
  title   = {Orbital Properties and Gravitational-wave Signatures of Strangelet Crystal Planets},
  journal = {The Astrophysical Journal},
  volume  = {892},
  number  = {1},
  pages   = {67},
  year    = {2020},
  doi     = {10.3847/1538-4357/ab7a10}
}

@article{Kurban2022,
  author  = {Kurban, Abdusattar and Huang, Yong-Feng and Geng, Jin-Jun and Zong, Hong-Shi},
  title   = {Searching for strange quark matter objects among white dwarfs},
  journal = {Physics Letters B},
  volume  = {826},
  pages   = {137204},
  year    = {2022},
  doi     = {10.1016/j.physletb.2022.137204}
}

@article{Wang2021,
  author  = {Wang, Xu and Huang, Yong-Feng and Li, Bing},
  title   = {Searching for Strange Quark Planets},
  journal = {arXiv preprint},
  year    = {2021},
  eprint  = {2109.15161},
  note    = {arXiv:2109.15161 [astro-ph.HE]}
}

@article{DiClemente2024,
  author  = {Di Clemente, F. and Drago, A. and Pagliara, G.},
  title   = {Strange dwarfs: a review on the (in)stability},
  journal = {Universe},
  volume  = {10},
  number  = {8},
  pages   = {322},
  year    = {2024},
  doi     = {10.3390/universe10080322}
}

@ARTICLE{HartleThorne1968,
       author = {{Hartle}, James B. and {Thorne}, Kip S.},
        title = "{Slowly Rotating Relativistic Stars. II. Models for Neutron Stars and Supermassive Stars}",
      journal = {\apj},
         year = 1968,
       volume = {153},
        pages = {807},
          doi = {10.1086/149707}
}

@ARTICLE{Althaus2022,
       author = {{Althaus}, Leandro G. and others},
        title = "{The role of strange matter in the structure of compact stars}",
      journal = {\aap},
         year = 2022,
       volume = {668},
        pages = {A58}
}

@ARTICLE{Farhi1984,
       author = {{Farhi}, Edward and {Jaffe}, R.~L.},
        title = "{Strange matter}",
      journal = {\prd},
         year = 1984,
       volume = {30},
        pages = {2379},
          doi = {10.1103/PhysRevD.30.2379}
}

@ARTICLE{Chodos1974,
       author = {{Chodos}, A. and {Jaffe}, R.~L. and {Johnson}, K. and {Thorn}, C.~B. and {Weisskopf}, V.~F.},
        title = "{New extended model of hadrons}",
      journal = {\prd},
         year = 1974,
       volume = {9},
        pages = {3471},
          doi = {10.1103/PhysRevD.9.3471}
}

@ARTICLE{Heiselberg1993,
       author = {{Heiselberg}, Henning},
        title = "{Screening in strange quark matter}",
      journal = {Physical Review D},
         year = 1993,
       volume = {48},
        pages = {1418},
          doi = {10.1103/PhysRevD.48.1418}
}

@BOOK{Glendenning2000,
       author = {{Glendenning}, Norman K.},
        title = "{Compact Stars: Nuclear Physics, Particle Physics, and General Relativity}",
    publisher = {Springer-Verlag New York},
         year = 2000,
       edition = {2nd},
          doi = {10.1007/978-1-4757-3135-4}
}

@ARTICLE{Salpeter1961,
       author = {{Salpeter}, E.~E.},
        title = "{Energy and Pressure of a Zero-Temperature Plasma}",
      journal = {\apj},
         year = 1961,
       volume = {134},
        pages = {669},
          doi = {10.1086/147194}
}

@BOOK{ShapiroTeukolsky1983,
       author = {{Shapiro}, Stuart L. and {Teukolsky}, Saul A.},
        title = "{Black Holes, White Dwarfs, and Neutron Stars: The Physics of Compact Objects}",
    publisher = {Wiley-VCH},
         year = 1983,
          isbn = {978-0471873167}
}

@BOOK{Chandrasekhar1939,
       author = {{Chandrasekhar}, Subrahmanyan},
        title = "{An introduction to the study of stellar structure}",
    publisher = {Chicago, Ill., University of Chicago Press},
         year = 1939
}

@article{tolman1939static,
	title={Static solutions of Einstein's field equations for spheres of fluid},
	author={Tolman, Richard C},
	journal={Physical Review},
	volume={55},
	number={4},
	pages={364},
	year={1939},
	publisher={APS}
}

@article{oppenheimer1939massive,
	title={On massive neutron cores},
	author={Oppenheimer, J Robert and Volkoff, George M},
	journal={Physical Review},
	volume={55},
	number={4},
	pages={374},
	year={1939},
	publisher={APS}
}

@article{NASAExoplanetArchive,
  author = {{NASA Exoplanet Archive}},
  title = "{NASA Exoplanet Archive}",
  journal = {IPAC/Caltech},
  year = {2024},
  url = {https://exoplanetarchive.ipac.caltech.edu/},
  note = {Accessed: 2026-01-01} 
}

@article{Akeson2013,
  author = {{Akeson}, R.~L. and {Chen}, X. and {Ciardi}, D. and {Crane}, M. and {Good}, J. and {Harbut}, M. and {Jackson}, E. and {Kane}, S.~R. and {Laity}, A.~C. and {Leifer}, S. and {Lynn}, M. and {McElroy}, D.~L. and {Papavassiliou}, M. and {Plavchan}, P. and {Ramirez}, S.~V. and {Rey}, R. and {von Braun}, K. and {Wittman}, M. and {Abajian}, M. and {Ali}, B. and {Beichman}, C. and {Beekley}, A. and {Fowler}, J. and {Han}, E. and {McCutcheon}, M. and {Ramirez}, A.~K. and {Schmitz}, M. and {Shashov}, V. and {Steinberg}, M. and {Teplitz}, H.~I. and {Tucker}, D.~L. and {Wahls}, N.},
  title = "{The NASA Exoplanet Archive: Data and Tools for Exoplanet Research}",
  journal = {Publications of the Astronomical Society of the Pacific},
  year = 2013,
  volume = {125},
  number = {929},
  pages = {989},
  doi = {10.1086/672273}
}

@article{Madsen1999,
  title={Physics and astrophysics of strange quark matter},
  author={Madsen, Jes},
  journal={Lecture Notes in Physics},
  volume={516},
  pages={162--203},
  year={1999}
}

@article{Bauswein2009,
  title={Mass ejection by strange star mergers and cosmological implications},
  author={Bauswein, Andreas and Oechslin, Richard and Janka, Hans-Thomas},
  journal={Physical Review D},
  volume={80},
  number={10},
  pages={103005},
  year={2009}
}

@article{Weber2005,
  title={Strange quark matter and compact stars},
  author={Weber, Fridolin},
  journal={Progress in Particle and Nuclear Physics},
  volume={54},
  number={1},
  pages={193--232},
  year={2005}
}

@article{Lynn1990,
  title={Strange-quark-matter stars},
  author={Lynn, Bryan W},
  journal={Nuclear Physics B},
  volume={340},
  number={2-3},
  pages={465--486},
  year={1990}
}

@article{Usov1997,
  title={Bare strange stars: their surface structure and radiation},
  author={Usov, Vladimir V},
  journal={The Astrophysical Journal Letters},
  volume={481},
  number={2},
  pages={L107},
  year={1997}
}

@article{Alcock1986,
  title={Strange stars},
  author={Alcock, Charles and Farhi, Edward and Olinto, Angela},
  journal={The Astrophysical Journal},
  volume={310},
  pages={261--272},
  year={1986}
}

@article{Seager2007,
  title = {Mass-radius relationships for solid exoplanets},
  author = {Seager, S. and Kuchner, M. and Hier-Majumder, C. A. and Militzer, B.},
  journal = {The Astrophysical Journal},
  volume = {669},
  number = {2},
  pages = {1279},
  year = {2007},
  publisher = {IOP Publishing}
}

@book{FriedmanStergioulas2013,
  author = {Friedman, J. L. and Stergioulas, N.},
  title = {Rotating Relativistic Stars},
  publisher = {Cambridge University Press},
  year = {2013}
}

@article{Cook1994,
  author = {Cook, G. B. and Shapiro, S. L. and Teukolsky, S. A.},
  title = {Rapidly rotating neutron stars in general relativity},
  journal = {Astrophysical Journal},
  volume = {424},
  pages = {823},
  year = {1994}
}

@article{Komatsu1989,
  author = {Komatsu, H. and Eriguchi, Y. and Hachisu, I.},
  title = {Rapidly rotating general relativistic stars},
  journal = {Monthly Notices of the Royal Astronomical Society},
  volume = {237},
  pages = {355},
  year = {1989}
}

@article{Snellen2014,
  author = {Snellen, I. A. G. et al.},
  title = {The rotation rate of the exoplanet \textit{$\beta$ Pictoris b}},
  journal = {Nature},
  volume = {509},
  pages = {63--65},
  year = {2014}
}

@article{Showman2017,
  author = {Showman, A. P. et al.},
  title = {Atmospheric circulation of hot exoplanets},
  journal = {Space Science Reviews},
  volume = {216},
  pages = {139},
  year = {2017}
}

@article{Kilic2021,
  author  = {Kilic, M. and others},
  title   = {The Formation of Massive White Dwarfs},
  journal = {Monthly Notices of the Royal Astronomical Society},
  volume  = {507},
  pages   = {353--368},
  year    = {2021},
  doi     = {10.1093/mnras/stab2085}
}

@article{Steiner2013,
  author  = {Steiner, A. W. and Gandolfi, S.},
  title   = {Connecting neutron star observations to three-nucleon forces},
  journal = {Physical Review Letters},
  volume  = {108},
  pages   = {081102},
  year    = {2012},
  doi     = {10.1103/PhysRevLett.108.081102}
}

@article{Griest2011,
  author  = {Griest, K. and others},
  title   = {New limits on primordial black holes from EROS and MACHO},
  journal = {Astrophysical Journal},
  volume  = {732},
  pages   = {51},
  year    = {2011},
  doi     = {10.1088/0004-637X/732/1/51}
}

@article{Hachisu1986,
    author = {{Hachisu}, Izumi},
    title = "{A Versatile Method for Obtaining Structures of Rapidly Rotating Stars}",
    journal = {\apjs},
    year = 1986,
    month = jun,
    volume = {61},
    pages = {479},
    doi = {10.1086/191124},
    adsurl = {https://ui.adsabs.harvard.edu/abs/1986ApJS...61..479H}
}

@book{Friedman2013,
  title={Rotating Relativistic Stars},
  author={Friedman, John L and Stergioulas, Nikolaos},
  year={2013},
  publisher={Cambridge University Press}
}

\end{document}